\documentclass[prd,prerprint,showpacs,showkeys,preprintnumbers,amsmath,amssymb,floatfix]{revtex4-2}
\usepackage{bm}
\usepackage{times,float}
\usepackage{graphicx}
\usepackage[T1]{fontenc}
\usepackage{bbold}
\usepackage{esint}
\usepackage[usenames,dvipsnames,svgnames]{xcolor}
\usepackage{hyperref}
\usepackage{multirow}
\usepackage{tabularx}
\usepackage{array}
\hypersetup{colorlinks=true, linkcolor=NavyBlue, citecolor=PineGreen,urlcolor=cyan}

\begin{document}	
\title{Numerical study of the properties of a holographic superconductor from an  anti-de
Sitter-Einstein-Born-Infeld black hole with backreaction}

\author{Uriel Elinos}
\email{elinosur@gmail.com}
\affiliation{Departamento de F\'isica, Universidad Aut\'onoma Metropolitana-Iztapalapa, Av. Ferrocarril San Rafael Atlixco 186, Ciudad de M\'exico 09310, M\'exico}
\author{Edgar Guzmán-González}
\email{edgar.guzman@hainanu.edu.cn}
\affiliation{School of Physics and Optoelectronic
Engineering, Hainan University, 570228 Haikou, P. R. China}
\affiliation{Departamento de F\'isica, Universidad Aut\'onoma Metropolitana-Iztapalapa, Av. Ferrocarril San Rafael Atlixco 186, Ciudad de M\'exico 09310, M\'exico}
\author{Marco Maceda}
\email{mmac@xanum.uam.mx}
\affiliation{Departamento de F\'isica, Universidad Aut\'onoma Metropolitana-Iztapalapa, Av. Ferrocarril San Rafael Atlixco 186, Ciudad de M\'exico 09310, M\'exico}

\date{\today}

\begin{abstract}
  We study numerically an $s$-wave holographic superconductor from an anti-de
  Sitter-Einstein-Born-Infeld black hole with backreaction in the context of the
  anti-de Sitter/conformal field theory (AdS/CFT) correspondence. By introducing a parameter $G$ to tune the effects of
  the backreaction, we can study non-perturbatively how the condensation and
  conductivity of the superconductor change as the backreaction increases. We
  find that for small values of $G$, increasing the nonlinearity of the
  Born-Infeld model makes the formation of the condensate harder --- consistent
  with previous results reported in the literature --- while for values of $G$
  close to one, the opposite effect occurs; increasing the nonlinearity slightly
  facilitates its formation. This result is confirmed by making similar studies
  for the electrical conductivity of the superconductor and the penetration
  depth of the Meissner effect. Moreover, we analize in detail the superfluid density in these systems. We also determine how the ratio $\omega_g/T_c$ varies for
  different intensities of nonlinearity and backreaction, showing in particular
  that large deviations from the so-called universal value arise as backreaction
  becomes stronger. 
\end{abstract}

\maketitle

\section{Introduction}

By this point in time, it is difficult to overstate the huge and fruitful impact that the AdS/CFT correspondence has on various areas of theoretical physics, like quantum information,
nuclear physics \cite{Guijosa2016,Natsuume2015,Edelstein2009} and condensed matter physics
\cite{Hartnoll2007,Sachdev2011,McGreevy2010,Hartnoll2009,Zaanen2015}. Said impact comes mainly from the fact that
it establishes a duality between a string theory and a conformal quantum field theory (CFT).
In certain cases, it is possible to work with models of string theory below 5 dimensions in an anti-de Sitter (AdS) background, within a sector that can be described as classical gravity. Even though these models have some limitations, they capture important phenomenological aspects of the dual system. 
In this setting, localized excitations in the CFT are described by data from the boundary on the gravity side; nevertheless a specific string theory is dual to a CFT at any depth of the bulk of the string theory. 

Many systems in condensed matter physics can be studied using this duality \cite{Ammon2010_pWaveHolographic}. An important example in this field that has been extensively studied using the correspondence is
the one of superconductors, as it is believed that they can provide the first steps to
understand theoretically the so-called high temperature superconductors \cite{Hartnoll2008_1, Hartnoll2008_Backreaction,Tong2013,Herzog2009}; these are strongly interacting systems that can not be described using the standard Bardeen-Cooper-Schrieffer (BCS) theory, as it presupposes
weak interactions \cite{Bardeen1957,Tinkham2004,Parks2018}. In this particular case, the duality is between a superconductor
and a black hole in the gravitational theory; for obvious reasons, the superconductors obtained from this correspondence are termed
holographic superconductors. 

One point of particular interest in this kind of analysis, is to determine how the properties of the
spacetime of the gravitational theory leave their imprint in the ones of the superconductor.
For this reason, different holographic superconductors have been studied previously in the literature, analyzing several
of their key properties like the expectation value of the superconducting condensate,
its critical temperature and its electrical conductivity.
Most of these studies follow a bottom-up approach, where one puts the basic ingredients of the theory of gravity by hand, without worrying about the details of how such theory can be embedded in supergravity.

There are at least two main objectives when building holographic superconductors: firstly, to study superconductors with strong interactions in a controlled way, and secondly, to understand how to change the parameters of the model
to engineer a superconductor with desirable properties, like having high critical temperature, high magnetic critical field or high values for the
condensate. The existence of excited states or more realistic scenarios where impurities are present may arise from models where modifications in the properties of spacetime are considered; among those that have been considered in the literature we may mention the inclusion of the Ricci curvature of the metric \cite{Gregory2009},
different nonrelativistic theories for the underlying gravity \cite{Taylor2008,Balasubramanian2008,Brynjolfsson2010,Momeni2011,Taylor2016},
nonlinear electrodynamic theories \cite{Pan2011,Zhao2013}, and noncommutative effects \cite{Ghorai2016,Pramanik2018, MacedaPatino2019}.

It is important to remark that when studying the gravitational theory, the resulting differential equations for the fields and the metric are nonlinear,
making an exact analytical treatment next to impossible, so one has to resort to approximations, numerical methods or a
combination of both.
Some examples of these types of techniques include the Sturm-Liouville method \cite{Siopsis2010,Wang2020} or
a matching method, where asymptotic solutions to the differential equations near the horizon of the black hole and near the boundary
are pasted consistently \cite{Gregory2009,Gangopadhyay2012, Roychowdhury2012}.
Due to the complexity of the resulting field equations, most of these works neglect the backreaction of the different matter fields on the metric,
so that the fields are on a fixed background.
Even though this approximation greatly simplifies the analysis and gives results that correctly capture the essential physics, 
to study temperatures considerably smaller than the critical one, or to obtain more precise quantitative results, numeric methods are necessary. 

In this work, we use the theoretical framework developed in \cite{Hartnoll2008_Backreaction} to numerically analyse an holographic superconductor with
backreaction, but instead of considering Maxwell electrodynamics, we consider Born-Infeld electrodynamics \cite{Born1934,Kerner2001}.
In Born-Infeld electrodynamics the self-energy of a point charge is regularized by introducing a maximum value $b$ for the magnitude
of the electric field. In the limits where either $b$ or the distance from the area of study to the sources goes to infinity,
Maxwell's theory is recovered. This makes holographic studies particularly simple, as the \emph{dictionary} used to translate
the properties of the CFT to the ones of the AdS are essentially the same as in the Maxwell case, as the boundary is far away from the sources.
Previous studies of holographic superconductors with Born-Infeld electrodynamics in the probe limit include \cite{Jing2010_BornInfeld,Kruglov2019}.
Studies beyond the probe limit were done for $(1+1)$ spacetimes \cite{Mohammadi2018},
for $(4+1)$ spacetimes \cite{Liu2012} and close to the probe limit \cite{Sheykhi2016,Ghorai_2018}.
Here we consider $(3+1)$ spacetimes further away from the probe limit and explore more in detail the formation of a condensate with a full backreacted metric.
Studies in the backreacted case in other gravitational backgrounds have also been carried out \cite{Brihaye2010_Backreaction,Liu2011_Backreaction,
 Momeni2013_AnalyticalBackreaction, Peng2013_StuckelbergBackreaction,Weiping2013, Chaturvedi2015} but employing techniques as the matching method and with a limited range of validity. As mentioned previously, with our numerical analysis, we gain insight into a totally backreacted Born-Infeld holographic superconductor; by its own nature, our approach covers non-perturbative aspects of holographic superconductors.

The structure of the paper is as follows,
in Section~\ref{sec:modelDefinition} we review the basic equations and formalism
to read the properties of the superconductor from the ones of the gravitational theory,
as well as how to set the initial conditions for the numerical methods.
We also discuss the effect of increasing the backreaction and the nonlinearity in the electrodynamical theory on
the condensation properties  and critical temperature of the holographic superconductor.
In Section \ref{sec:conductivity} we make a similar study to the previous one but
for AC conductivity of the superconductor near the critical temperature; we discuss there the behaviour of the ratio $\omega_g/T_c$ as function of the Born-Infeld parameter and backreaction. Afterwards, we analyse the consequences of backreaction on the Meissner effect using the formalism of mixed boundary conditions in Section~\ref{sec:meissner}.
Finally, in Section~\ref{sec:summary} we present our general conclusions, a summary of our results and some possible directions for further research.

\section{Model for the holographic superconductor}
\label{sec:modelDefinition}
First, we briefly describe the bulk theory that we are considering. The dynamics of the model, in natural units, is given by the following
action (assuming a sum over repeated indices running from 0 to 3)
\begin{equation}
  \begin{split}
    S &= \int d^{3+1} x\, \sqrt{-g}\,\mathcal L,
        \\
 \mathcal L&=
    \frac{1}{16 \pi} \Big(G^{-1}(R - 2 \Lambda)+\mathcal L_{BI}\Big) - \frac{m^{2}}{L^{2}} \psi \psi^{*}
   \\ &{}- (\partial_{\alpha} \psi - i q A_{\alpha} \psi)g^{\alpha \beta} (\partial_{\beta} \psi^{*} + i q A_{\beta} \psi^{*}),
  \end{split}
  \label{eq:action}
\end{equation}
where  $g$ is the metric, $R$ is its Ricci scalar, $L$ is the AdS radius and $\Lambda$ is a cosmological constant;
 $\psi$ is a complex scalar field with charge $q$ and mass $m$.
$A_\mu$ is the electromagnetic potential associated with the Born-Infeld Lagrangian $\mathcal L_{BI}$
defined in terms of the electromagnetic tensor $F_{\alpha \beta} \equiv \partial_{\alpha} A_{\beta}- \partial_\beta A_{\alpha}$
as follows,
\begin{equation}
  \begin{split}
\mathcal L_{BI}=
4 b^2 \left(1-\sqrt{1+\frac{F_{\alpha \beta} F^{\alpha \beta} }{2 b^2}}\right),
  \end{split}
\end{equation}
with $b$ the Born-Infeld parameter, that physically is the maximum value
for the magnitude of the electric field, and
$G$ is the universal gravitational constant.
Clearly, at regions where $| F_{\alpha \beta}| \ll b$, we recover Maxwell's linear theory,  $L_{BI}(F) \approx - F^{\alpha \beta} F_{\alpha \beta}$.

Note that, as done in \cite{Ammon2010_pWaveHolographic}, $G$ can be regarded as a parameter that controls the effects of the backreaction
--- when $G=0$ the dynamics of the metric is given only by Einstein's term $R-2 \Lambda$;
as $G$ keeps increasing, the effects on the metric of the other fields also increase
and $G=1$ defines our \emph{maximal} backreacted case.

The variation of $S$ with respect to the fields $g$, $\psi$, $\psi^{*}$  and $A_\mu$ produces all the equations
of motion for the system. This theory allows us to have a system that admits hairy solutions at low temperatures and not at high ones and is dual to a CFT that has a conserved current and a charged operator to break the $U(1)$ symmetry. Also, in order to obtain a theory that is dual to an $s$-wave superconductor on the CFT, we may consider a plane-symmetric metric
(other options for the metric's symmetry include spherical or hyperbolic, but all of them are related to different radial foliations associated with the metric where the CFT resides, resulting in different approaches to duality with similar results to the ones presented here \cite{Guijosa2016})
with a charged black hole (say, at the origin) that is asymptotically  anti-de Sitter  \cite{Hartnoll2008_1}. The simplest metric
that satisfies these requirements is of the following form,
 \begin{equation}
   \begin{split}
ds^2 = -f(r) dt^{2} + \frac{d r^{2}}{f(r) n(r)} + r^{2} (dx^{2}+dy^{2}),
   \end{split}
   \label{eq:metricAnsatz}
 \end{equation}
where $t,r,x,y,$ are coordinates of the $4D$ spacetime, being $r$ a radial coordinate,
$f(r)$ is the metric function and $n(r)$ measures  the effects of the backreaction --- indeed,
$n(r)\equiv 1$ in the probe limit. The boundary of the AdS, where the CFT {``lives''}, is located at $r \rightarrow \infty$.
The value $r_{+}$, where $f$ vanishes, is the horizon radius of the black hole. The temperature of the black hole is given by
Hawking's celebrated expression (we assume that $n(r_{+}) \neq 0$ and that $f(r), n(r)$ are positive for $r>r_{+}$),
\begin{equation}
  \begin{split}
T = \frac{\sqrt{n(r_{+})}f'(r_{+}) }{4 \pi }.
  \end{split}
  \label{eq:hawkingTempBackReaction}
\end{equation}
Since we have a plane-symmetric metric, we
assume that $\psi$ only depends on $r$, and, since we will assume that there is no magnetic field,
we initially work on a gauge where the potential takes the following simple form,
$A_{t}= \phi(r)$, $A_{r}=A_{x}=A_{y}=0$.
In terms of this \emph{Ansatz}, Einstein's equations for the components $tt$ and $rr$ become
(where a prime ${}'$ denotes the derivative with respect to $r$),
\begin{equation}
  \begin{split}
0=&16 \pi G q^2 r \psi \psi^* \phi ^2+f^2 (n'+16 \pi  G r n \psi ' \psi^{*}{}'),\\
0=&
\frac{4 b^2 r^2 G}{\sqrt{1-\frac{n \phi'{}^2}{b^2}}}-2 n (r f'+f)-r f n'
\\&{}-16 \pi  G m^2 r^{2} \psi \psi^{*}-2 \Lambda r^{2} +4 b^2 r^{2} G,
  \end{split}
  \label{eq:f-n-equations}
\end{equation}
while  Born-Infeld's equation  for $A_{t}$ and
the one for the scalar field $\psi$ become,
\begin{equation}
  \begin{split}
0=&
8 \pi  q^2 r^2 \psi \psi^{*}  \phi -f \sqrt{n} \Bigg(
\frac{r^2 \sqrt{n} \phi '}{\sqrt{1-\frac{n \phi '{}^2}{b^2}}}
\Bigg)',
  \\
0=&
   \left(q^2 \phi ^2-m^2 f\right)r^{2}\psi
+{f \sqrt{n}} \Big(r^2 f \sqrt{n} \psi ' \Big )'.
  \end{split}
  \label{eq:phi-psi-equations}
\end{equation}
The remaining Einstein's and Born-Infeld's equations can be proved to be a combination of these four,
so they are satisfied automatically if Eqs.~(\ref{eq:f-n-equations}) and (\ref{eq:phi-psi-equations}) are.
It is also clear from Eq.~(\ref{eq:phi-psi-equations}) that,
since we can assume real initial conditions for $\psi$ (because of the invariance of the action in Eq.~(\ref{eq:action})  under the transformation
$\psi \rightarrow  e^{i \gamma}\psi$, with $\gamma$ real), we can assume that $\psi$ is real throughout all the solution.

Once we have obtained the equations of motion, our goal is to solve them in full generality for arbitrary values of the Born-Infeld parameter; moreover, we want to analyse the behaviour of the solutions when backreaction becomes important. For this purpose, we have to make use of numerical solutions that allow us to explore the full moduli space of the model; in this way, we will obtain a complete characterization of the associated holographic superconductor, in contrast to previous treatments in the literature where effects linear on the Born-Infeld parameter have been discussed and where backreaction has also been considered perturbatively.

The approach we use to solve the previous system of equations numerically is the \emph{shooting method}. As a first step towards its implementation, we fix the parameters  of the model to the values $L=1$, $q=1$, $m^2=-2$. Next, we impose some regularity and physical conditions at the horizon $r_{+}$ and at the boundary; specifically, we begin discussing the asymptotics near the boundary since several interesting observables of the CFT can be read from the asymptotic behaviour of the fields at the boundary,
as we remind the reader below.

By studying the equations of motion when $r \rightarrow \infty$, we see that the fields have the following asymptotic form
\begin{equation}
  \begin{split}
\psi(r) &= \frac{\mathcal J_{O}}{r}+ \frac{\langle O_{2} \rangle}{\sqrt{2}r^{2}} + \dots\,,
\\
\phi(r) &= \mu- \frac{\rho}{r} + \dots\,,
\\
f(r) &= \kappa^{-1}r^{2}+ f^{0}+ \frac{f^{1}}{r} + \dots\,,
\\
n(r) &= \kappa + \frac{n_\infty^{1}}{r} +\dots\,.
  \end{split}
  \label{eq:psiphiInfinityExpansion}
\end{equation}
where, according to the AdS/CFT dictionary \cite{Klebanov1999,Hartnoll2008_1,Zaanen2015}, with the gauge we are considering,
$\langle O_{2} \rangle$ is the expectation value of the superconducting condensate,
$\mathcal J_{O}$, the one of its source, whilst $\mu$ is the chemical potential and
 $\rho$ is the charge density of the field theory. $f^{0}$ is related to $\mathcal J_{O}$ and $f^{1}$ is related to the free energy of the system; the
particular relationship will not be of interest for our work.
To study unsourced spontaneous condensation, we set $\mathcal J_{O} = 0$. This is a condition that has to
be implemented in the shooting method.

In order for the temperature of the black hole in Eq.~(\ref{eq:hawkingTempBackReaction}) to coincide with the one of the CFT, it is required
that $\kappa=1$ \cite{Hartnoll2008_Backreaction}. In principle, this would be another condition for the shooting method,
however, we can simplify this step by
fixing $n(r_{+})$ to an arbitrary value (say, $n(r_{+})=1$), solve for the remaining conditions using the shooting method
and then apply the following scaling symmetry of the field equations that map solutions of the equations of motion
to solutions (note that this scaling is the one induced by the coordinate transformation $t' = \kappa t$),
\begin{equation}
  \begin{split}
f \rightarrow \kappa  f,
\, n\rightarrow  n / \kappa ,
\,\phi\rightarrow \sqrt{\kappa}  \phi,
\,\psi\rightarrow \psi.
  \end{split}
  \label{eq:scalingFields}
\end{equation}
As can be easily verified, this scaling 
satisfies
the remaining conditions $\mathcal J_{O}=0$ and Eq.~(\ref{eq:conditions-rh}) down below that we enforce on the shooting method.
By imposing regularity of the fields in Eq.~(\ref{eq:psiphiInfinityExpansion}), the value $\Lambda=-3$ also gets fixed.

To ensure a divergence-free QFT dual to a charged black hole, it is necessary that the fields $f$, $n$, $\Psi$, and $\phi$ are analytic at the horizon \cite{Musso2014}. Thus, we propose Taylor expansions for these fields around $r=r_+$ and find the relationships among the coefficients of the expansion that ensure their analyticity. These relationships are also important for the numerical implementation of the shooting method. Since $f(r_+)=0$, the ordinary differential equations in Eq. (\ref{eq:phi-psi-equations}) are singular at $r_+$, so we cannot shoot solutions from $r_+$. To circumvent this limitation, we utilize the obtained expansions to evaluate the fields and their derivatives at $r_++\epsilon$, with $\epsilon \ll r_+$, providing the necessary initial conditions to shoot solutions from $r_++\epsilon$, where the system is not singular.
Therefore, if we assume
an expansion of the form,
\begin{equation}
  \begin{split}
f(r) &= f_{1}(r-r_{+})+ f_{2}(r-r_{+})^{2}+ \dots\, , \\
n(r) &= n_{0}+n_{1}(r-r_{+})+ n_{2}(r-r_{+})^{2}+ \dots\, , \\
\psi(r) &= \psi_{0}+\psi_{1}(r-r_{+})+ \psi_{2}(r-r_{+})^{2}+ \dots\, , \\
\phi(r) &= \phi_{0}+\phi_{1}(r-r_{+})+ \phi_{2}(r-r_{+})^{2}+ \dots\, ,
  \end{split}
\end{equation}
and impose analyticity, we conclude the following relations,
\begin{equation}
  \begin{split}
    f_{1}&=
\frac{r_{+} }{n_{0}}\left(2 b^2 G-\frac{2 b^2 G}{\sqrt{1-\frac{n_0 \phi_1^2}{b^2}}}-8 \pi  G m^2 \psi_0^2
-\Lambda \right)\, , \\
    \phi_{0}&=0\, , \quad \psi_{1}= \frac{m^2 \psi_{0} }{f_{1} n_{0}}\, .
  \end{split}
  \label{eq:conditions-rh}
\end{equation}
As already mentioned, we can set $n_{0}=1$ and then apply a scaling if necessary. The previous equations fix the remaining
necessary conditions in terms of just $\psi_0$ and $\phi_1$ (fixing $b$, $q$, $G$ and $m$), making these two the only parameters we need to vary when applying the
shooting method.

So in summary, to apply the shooting method, we do the following: First, we choose a value for $r_{+}$, in our case,
we set $r_{+}=1$. Then, we solve the system of Eqs.~(\ref{eq:f-n-equations}), (\ref{eq:phi-psi-equations}), (\ref{eq:conditions-rh}), $f(r_{+})=0$ and $n_{0}=1$, keeping $\psi_0$ and $\phi_1$ as free parameters.
Once we have solved the system of equations numerically for some pair $(\psi_0,\phi_1)$,
we can do a fit for $r \gg r_{+}$ to compute
$\mathcal J_{O}$ and the remaining observables of Eq.~(\ref{eq:psiphiInfinityExpansion}). By varying the values of
$\psi_{0}$ and $\phi_{1}$,
we can find a curve in the $\psi_{0}$-$\phi_{1}$ plane such that $\mathcal J_{O}=0$. After applying the scaling in Eq.~(\ref{eq:scalingFields})
to the fields obtained by this procedure, we can use their data to study different properties of the unsourced superconducting condensate.

\begin{figure}
	\includegraphics[scale=0.7]{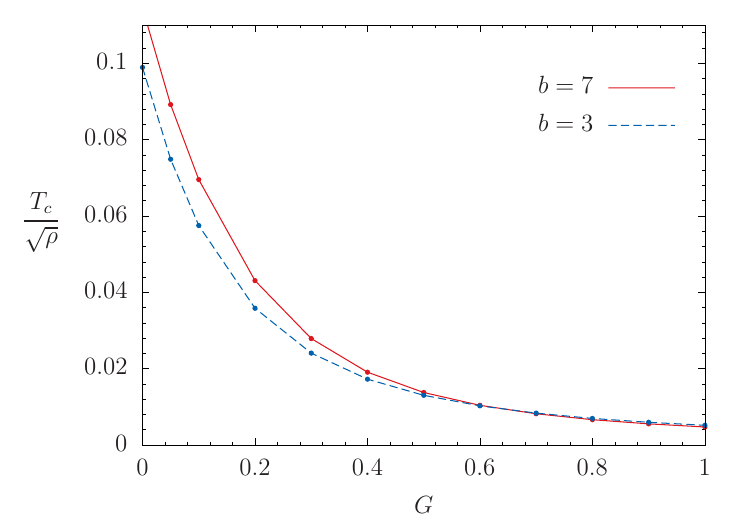}
	\caption{Plot of the ratio $T_{c}/\sqrt{\rho}$ as a function of $G$ with $q=1$ and different values of $b$.
From the left side, the upper curve is for $b=7$ while the lower one is for $b=3$. Notice how, for $G$ close to
one, the situation reverses, the $b=3$ curve is now above the $b=7$ one, indicating that the formation of the condensate
 is easier for $b=3$.}
	\label{fig:G-Tc-rho}
\end{figure}

Since in a QFT only dimensionless quantities have physical meaning, we focus on the ratio $T/\sqrt{\rho}$
(we could consider other dimensionless ratios like $T/ \mu$, but they all produce similar results).
The parameters $\psi_{0}$, $\phi_1$, where
$\langle O_{2} \rangle$ becomes zero, are the ones that correspond to the solution related to the critical temperature $T_{c}$; by finding them numerically,
we can compute the ratio $T_c/ \sqrt{\rho}$ as a function of $G$ and $b$. In Fig.~\ref{fig:G-Tc-rho}, we fix the values $b=7$, $b=3$ and
plot $T_{c}/\sqrt{\rho}$ as a function of $G$. Note that, when $G$ increases, the ratio $T_{c}/\sqrt{\rho}$ becomes smaller;
this means that, at fixed $\rho$, increasing $G$ implies a lower critical temperature (i.e.\ the formation of the super conducting
condensate becomes harder). For low values of $G$, we see a similar behaviour for $b$, since the $b=7$ curve is above
the $b=3$ one; we see that decreasing the values of $b$ (increasing the
nonlinearity on the electrodynamic theory) makes the condensation harder; however, there is a value $G^{*}$
(somewhere between $G=0.6$ and $G=0.7$) such that, for $G > G^*$, increasing the nonlinearity of the electrodynamical theory
makes the condensation slightly easier.

\begin{figure}
	\centering
	\includegraphics[scale=0.7]{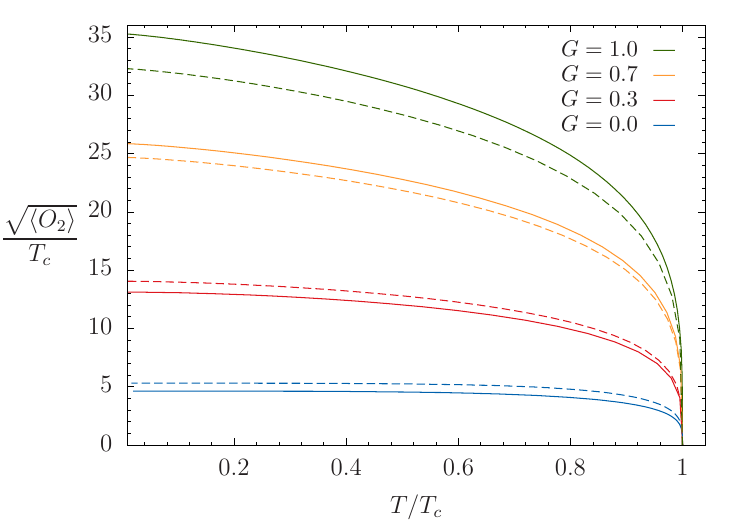}
	\caption{Plot of $T/T_{c}$ {\it vs} $\sqrt{\langle O_{2} \rangle}/T_{c}$ for different values of $G$. The dashed curves correspond to $b=3$, while
continuous lines are for $b=7$. The values of $G$ considered are, from top to bottom, $G=1,0.7,0.3,0$. Notice how, for
$G=0,0.3$ (small backreaction), the $b=7$ curves are below the $b=3$ curves, indicating that the formation of the condensate
is harder for $b=3$, however, for $G=0.7,1$ the situation reverses, condensation becomes easier for $b=3$.}
	\label{fig:T-Tc-O2-Tc}
\end{figure}

With the proportionality constant between $T_{c}$ and $\rho$ determined, we can obtain similar conclusions to the previous ones
by considering  plots of $T/T_{c}$ {\it vs} $\sqrt{\langle O_{2} \rangle}/T_{c}$ for different values of $b$ and $G$. The results are presented in Fig.~\ref{fig:T-Tc-O2-Tc}. In these types of curves, the higher the values of $\sqrt{\langle O_{2} \rangle}/T_{c}$, the harder the formation
of superconducting condensate; this behaviour can be properly explained by noticing that the value of the critical temperature $T_{c}$ gets lower and this indicates that the superconducting state has been destabilized. This also can be analyzed noticing that the value of the condensate is proportional to the energy gap of the superconducting state, and from these results we can see that for a fixed gap energy, increasing the backreaction requires lower critical temperature to achieve it. From here we confirm the previous conclusions, for $G \in \{0,0.3\}$, condensation is easier
for $b=7$, while, for $G\in \{0.7,1.0\}$ the condensation is easier for $b=3$. We also mention that, as found in previous works,
\cite{Gangopadhyay2012, Chaturvedi2015},
from our numerical analysis we observe the critical exponent of 1/2 characteristic of second order phase transitions,
$\langle O_{2} \rangle \propto (1-T/T_{c})^{1/2}$, independently of $b$ and $G$.

\section{Conductivity}
\label{sec:conductivity}
With the AdS/CFT correspondence, we can study the AC conductivity of the superconducting condensate in the usual way ---
we turn on a small harmonic electric field with frequency $\omega$ and study the linear response of the system \cite{Son2002}.
In our setting, the simplest electric field is one that has only either the $x$ or $y$ component --- for concreteness we chose the $x$
component--- and zero momentum (implying that there will be no heat flow in the superconductor \cite{Hartnoll2008_Backreaction}),
so we set $A_{x}(r,t)= \epsilon e^{- i \omega t} A_{x}(r)$, and study the system up to first order in $\epsilon$.

With $A_{x} \neq 0$, the \emph{Ansatz} for the metric in Eq.~(\ref{eq:metricAnsatz}) is no longer consistent, as Einstein's equation for
the $tx$ component is independent from the remaining equations, producing a system with more independent equations that functions and
hence, with no solutions generically.
To the first order in $\epsilon$, this can be easily
solved by adding the term $2\epsilon g_{xt}(r) e^{-i \omega t} dt dx$ to the \emph{Ansatz} in Eq.~(\ref{eq:metricAnsatz}).
To this order, Eqs.~(\ref{eq:phi-psi-equations}) and  (\ref{eq:psiphiInfinityExpansion}) for the original fields
remain the same, while we find the following new independent equations,
\begin{equation}
  \begin{split}
0&=\frac{4 G {A_x} \phi '}{\sqrt{1-\frac{n \phi '{}^2}{b^2}}}+g_{xt}'-\frac{2 g_{xt}}{r}\,,\\
0&=
   \Bigg(
\frac{\omega ^2}{f n}
-\frac{4 G \phi '{}^2}{\sqrt{1-\frac{n \phi '{}^2}{b^2}}}-\frac{8 \pi  q^2 \psi ^2 \sqrt{1-\frac{n \phi '{}^2}{b^2}}}{n}
   \Bigg)
\frac{A_{x}}{f} \\
   &{}+
   \Bigg(
\frac{n'}{2n}+\frac{f'}{f}+\frac{8 \pi  q^2 \psi ^2 \phi  \phi ' \sqrt{1-\frac{n \phi '{}^2}{b^2}}}{b^2f}-\frac{2 n \phi '{}^2}{b^2r}
   \Bigg)
A_{x}'
     \\&{}+A_{x}'',
  \end{split}
  \label{eq:Ax}
\end{equation}
where we have written $g_{xt}$ in terms of $A_x$ and its derivatives in the last equation,
so it is not necessary to find $g_{xt}$ to determine the remaining fields. Therefore, it is enough
to just discuss the conditions for $A_{x}$ at $r_{+}$ and at the boundary.

At $r_{+}$, we require that $A_{x}$ describes an ingoing wave (reflecting the fact that no electromagnetic wave
gets out of a black hole \cite{Musso2014}), so we assume an expansion near $r_{+}$ of the form,
\begin{equation}
  \begin{split}
A_{x}(r)=(r - r_{+})^{-i \kappa} (A_{x0}+A_{x1}(r - r_{+})+ \dots)\,,
  \end{split}
\end{equation}
with $\kappa >0$. By considering this expansion in Eq.~(\ref{eq:Ax}) we conclude,
\begin{equation}
  \begin{split}
\kappa = \frac{\omega}{{f_1 \sqrt{n_0}}},
  \end{split}
\end{equation}
together with a linear relation between $A_{x0}$ and $A_{x1}$ that we do not present here because of its considerably big size.
From this analysis we conclude that we only have to specify a single condition for $A_{x}$; since Eq.~(\ref{eq:Ax}) {is}
linear in $A_{x}$ by construction, and we are interested in the ratio of two coefficients related to $A_{x}$ (see Eq.~(\ref{eq:AxBoundary})),
we can simply set $A_{x0}=1$.

Furthermore, by considering Eq.~(\ref{eq:Ax}) once more, we see that at the boundary, $A_{x}$ has the following asymptotic expression,
\begin{equation}
  \begin{split}
A_{x}(r)= -\frac{i \mathcal E}{\omega}+\frac{\mathcal J_{\mathcal E}}{r}+ \dots,
  \end{split}
  \label{eq:AxBoundary}
\end{equation}
where $\mathcal E$ is the $x$ component in the {dual} theory of the electric field produced by
the time dependent electromagnetic potential and $\mathcal J_{\mathcal E}$ is the corresponding current.
The previous expression can be obtained either by naive considerations or by using the GKPW prescription \cite{Gubser1998}.
Although the latter computation for the Born-Infeld's case is similar to the one of Maxwell's \cite{Hartnoll2008_Backreaction},
it is instructive to derive it, so we include it in Appendix \ref{appendix:A}.

Either way, by using Ohm's law, we can directly compute the
conductivity $\sigma$ from  $\mathcal J_{\mathcal E}=\sigma  \mathcal E  $ \cite{Siopsis2010}.
Similar techniques can be used to compute other transport properties, like the heat current \cite{Musso2014,Donos2017} or the
correlation length \cite{Maeda2009}.

\begin{figure}
	\centering
	\includegraphics[scale=0.7]{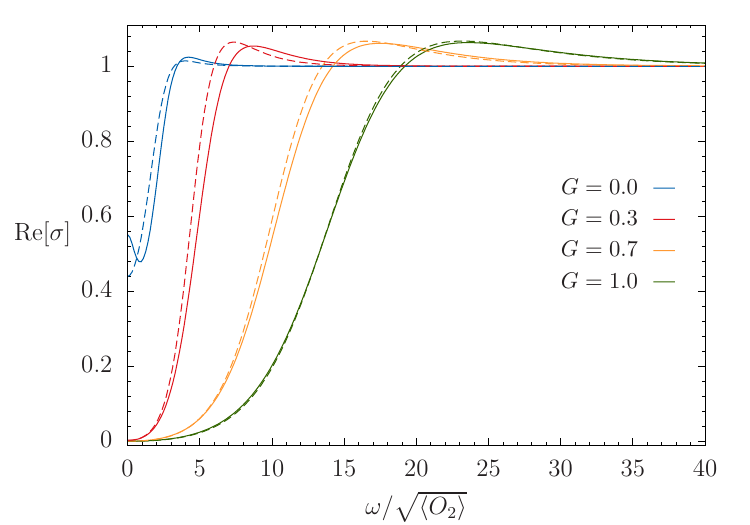}
	\caption{Plot of $\omega/\sqrt{\langle O_{2} \rangle}$ {\it vs} $\mathrm{Re} [\sigma]$ for different values of $b$ and $G$ at a fixed ratio $T/T_{c}=0.9$.
   Dashed curves are for $b=3$, while the continuous one are for $b=7$. The values of $G$ considered are, from left to right, $G=0,0.3,0.7,1.0$. Notice
   how $\omega_{g}$, the frequency where $\mathrm{Re} [\sigma]$ grows abruptly, is very sensible to the value of $G$.}
	\label{fig:omega-reSigma}
\end{figure}

In Fig.~\ref{fig:omega-reSigma}, we present a plot of the frequency $\omega$ (measured in units of $\langle\mathcal O_2\rangle$) \textit{vs.} the real part of the conductivity $\mathrm{Re}[\sigma]$ for a fixed temperature $T=0.9 T_c$ and different values of $b$ and $G$. The behavior of $\mathrm{Re}[\sigma]$ can be explained by describing the superconductor in terms of the two-fluid model and studying the response of the ``superconducting'' and the ``normal'' electrons to alternating electromagnetic fields. According to BCS theory, at $T=0$, there are no normal electrons for small $\omega$, and for $T>0$, their presence is related to thermal excitations \cite{Tinkham2004}. Precisely at $T=0$, the only mechanism available for the system to absorb an incident electromagnetic wave is by creating pairs of electrons. Therefore, for frequencies below the gap frequency $\omega_g$---the frequency associated with the necessary energy to break a Cooper pair into a pair of electrons \cite{Musso2014}---$\mathrm{Re}[\sigma]$ is zero \cite{Mattis1958,Chen1993}. At $\omega_g$, there is an absorption edge, and the conductivity quickly increases from zero to unity, resembling the behavior of a holographic conductor in the normal state \cite{Herzog2007}. For $T>0$, although most of the current is supercurrent, the nonzero $\mathrm{Re}[\sigma]$ is related to the presence of normal electrons mentioned earlier. In this case, $\omega_{g}$ can be detected as the frequency where $\mathrm{Re}[\sigma]$ starts growing abruptly, and it coincides with the frequency where $\mathrm{Im}[\sigma]$ attains a minimum value \cite{Horowitz2008}.
With this in mind, we can observe in Fig.~\ref{fig:omega-reSigma} how the energy gap increases as we increase $G$, while it remains almost unchanged when we vary from $b=3$ to $b=7$. This can be explained 
by recalling that higher values of $G$ imply lower critical temperatures, so the energy necessary to break a Cooper pair increases.

\begin{figure}
	\centering
	\includegraphics[scale=0.7]{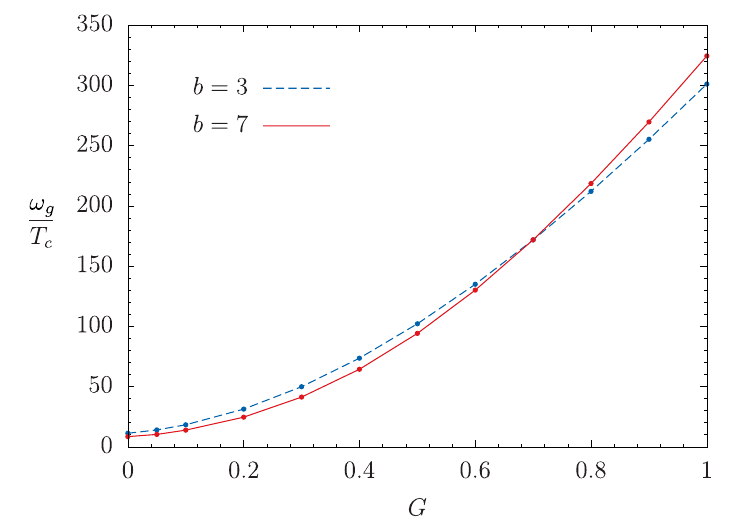}
	\caption{Plot of the ratio $\omega_{g}/T_{c}$ for different values of $G$ and $b$, at fixed ratio $T/T_{c}=0.9$. Dashed curves are for $b=3$,
   and continous curves, for $b=7$. Large deviations from the universal value $\omega_g/T_{c}\approx8$ seen in the Einstein-Maxwell case are observed.
   Notice that, like in previous figures, for $G$ close to zero, the $b=3$ curve is above the $b=7$ one,
   but at a certain $G$ between $0.6$ and $0.7$, the situation reverses.}
	\label{fig:G-omega}
\end{figure}

In \cite{Horowitz2008}, it was seen that for the Einstein-Maxwell case, the universal value $\omega_{g}/T_{c}\approx 8$ holds for different values of
the remaining parameters. It also holds in other settings, even when noncommutative effects of the underlying spacetime are present
\cite{MacedaPatino2019}. However, it was seen that when electrodynamics or gravitational theories change,
the value of this ratio can change significantly \cite{Gregory2009,Sheykhi2018}. In Fig.~\ref{fig:G-omega} we plot
it for different values of $G$ and $b$. We see that for strong backreaction, the ratio can vary considerably from $8$.
Nonetheless, the ratio is still different from the value $\omega_g /T_{c} \approx 3.5$ predicted by the BCS theory \cite{Tinkham2004,Herzog2009},
confirming that this is still a superconductor with strong interactions.

\section{Meissner effect}
\label{sec:meissner}

Finally, we investigate the response of the holographic superconductor to weak
magnetic fields. Particularly, we study the Meissner effect by computing the
penetration depth of a small magnetic field, as recently discussed
in \cite{Natsuume:2022kic,Krikun:2023ijs}. To achieve this, we apply a perturbation
$A_x(u,y)$ to the $x$ component of the gauge field, where $u \equiv r_+ / r=1/r$, and
study the system's response. A key aspect
of this approach is to promote the boundary gauge field to a
dynamical one by imposing ``mixed'' boundary conditions given by the semiclassical
Maxwell equations $\partial_{j} \mathcal{F}^{ij}=4 \pi \mathcal{J}^{i}$, where $\mathcal{F}^{ij}$ denotes the
boundary electromagnetic tensor, and  $\mathcal{J}^{i}$ represents the corresponding boundary current.

As a warm-up and to study the different contributions of the backreaction, we
first adopt a hybrid approach valid near the probe limit, where we initially neglect the
backreaction on the metric when calculating the penetration depth $\lambda$ and incorporate some of its effects afterwards.
Specifically, while computing $\lambda$, we assume the metric to be
that of a Schwarzschild-AdS$_4$ black hole,
\begin{equation}
ds^{2} = \left( \frac{r_{0}}{u} \right) ^{2} (-f dt^{2} + dx^{2} + dy^{2})+ \frac{du^{2}}{u^{2} f}, \quad
f =  1-u^{3}.
\end{equation}
However, in the resulting expression for the penetration depth, we use the
fields $\phi$ and $\psi$ that were obtained by numerically solving the equations for the backreacted case, namely Eqs.~(\ref{eq:f-n-equations}) and (\ref{eq:phi-psi-equations}). At the end of this section, we study the fully backreacted scenario,
where the backreaction on the metric is taken into account throughout the computation.

A straightforward calculation shows that in the hybrid case, to the first order, the perturbation $A_x$ must satisfy the following equation,
\begin{align}
\label{eq:meissnerperturb}
& \partial_{u} \left(\frac{f}{\xi} \partial_{u} A_{x}\right) + \frac{1}{\xi} \partial_{y}^{2} A_{x} - \frac{8 \pi q^{2} A_{x} \psi^{2}}{u^{2}}=0, \qquad \xi (u)\equiv \sqrt{1- \frac{u^{4} \phi'^{2}}{b^{2}}}.
\end{align}
By imposing the mixed boundary conditions $\partial_{j} \mathcal F^{ij}= 4 \pi \mathcal{J}^{i}$ and doing an expansion to the second order in $\psi$,
we can compute $\lambda$ \cite{Natsuume:2022kic} (see Appendix \ref{appendix:B} for the detailed computation),
\begin{equation}
    \lambda ^{2}= \frac{1}{32 \pi^2 q^{2} \mu_{m} I}, \qquad I\equiv \int_{0}^{1} du \frac{\psi^{2} (u)}{u^{2}},
    \label{eq:lambdaMeissnerProbe}
\end{equation}
where $\mu_m$ is the magnetic permeability from the normal phase, which is given by,
\begin{equation}
\mu_{m}= \frac{1}{1+ 4 \pi \beta}, \qquad \beta\equiv \int_{0}^{1} du \frac{1}{\xi (u)}.
\label{eq:mubetaIprobe}
\end{equation}

From these calculations, we shall see that $\lambda^2$ behaves as
expected: the lower the temperature, the shorter the penetration depth,
indicating that the magnetic field is expelled more effectively and decays
more rapidly inside the bulk of the superconductor. As
$T$ approaches the critical temperature $T_c$, $\lambda^2$ diverges, showing that the magnetic field penetrates
deeper. Even in this case, where certain contributions of the backreaction are
neglected, we observe that higher values of $G$ increase $\lambda$, indicating that
the backreaction tends to weaken the superconducting properties.

The next step is to consider the fully backreacted scenario, where the effects of $A_x$ and the remaining fields
on the metric are taken into account. As was the case when computing the conductivity in
Section \ref{sec:conductivity}, the \emph{Ansatz}  for the metric in Eq. (\ref{eq:metricAnsatz}) becomes
inconsistent in the presence of the additional component of the gauge field,
requiring a non-zero
$g_{tx}(u,y)$
component for consistency. To first order in $A_x$ and
$g_{tx}$, this leads to the following system of equations,
\begin{align}
 \sqrt{n} u^{2} \partial_u\left( \frac{f \sqrt{n} u^{2} \partial_u{A}_{x} + \sqrt{n} u^{2} {g}_{tx} \partial_u \phi }{4 \pi \gamma} \right) - 2 q^{2} \psi^{2} {A}_{x} - \frac{2 q^{2} \psi^{2} \phi {g}_{tx}}{f} + \frac{ \partial_y^2 {A}_{x} u^{2}}{4 \pi \gamma} &=0 \, , \\
 \sqrt{n} u^{2} \partial_u \left( \sqrt{n} u^{2} \partial_u {g}_{tx} \right) + \frac{4 G n u^{4} \partial_u \phi\, \partial_u {A}_{x}}{\gamma} + \frac{32 \pi G q^{2} \psi^{2} \phi {A}_{x}}{f} - \frac{u^{2}}{f} \left( 2fn{g}_{tx} - \partial_y^{2}{g}_{tx} - u f{g}_{tx} \partial_u n  \right) &=0\, ,
\end{align}
where $\gamma\equiv \sqrt{1- u^{4} n \phi'^{2}/b^{2}}$.
The procedure for computing $\lambda$ in this case is similar to the previous one, but
there are additional details to consider. The complete calculation is provided
in Appendix \ref{appendix:C}. In the specific case in which there are no sources of heat flow at the boundary,
the resulting expression for $\lambda$ remains the same as in Eq.~(\ref{eq:lambdaMeissnerProbe}), but
$\mu_m$, $\beta$ and $I$ are modified as follows,
\begin{equation}
\mu_{m}= \frac{1}{1+ 4 \pi \beta}, \qquad \beta= \int_{0}^{1} du \frac{1}{\sqrt{n(u)}\gamma (u)}, \qquad I= \int_{0}^{1} du \frac{\psi^{2} (u)}{u^{2} \sqrt{n(u)}}.
\label{eq:mubetaIbackreaction}
\end{equation}
The expression for $\lambda$ when there is a source of heat flow is provided in Eq.~(\ref{eq:lamba2Final}) of Appendix \ref{appendix:C}. 

\begin{figure}
	\centering
	\includegraphics[scale=0.75]{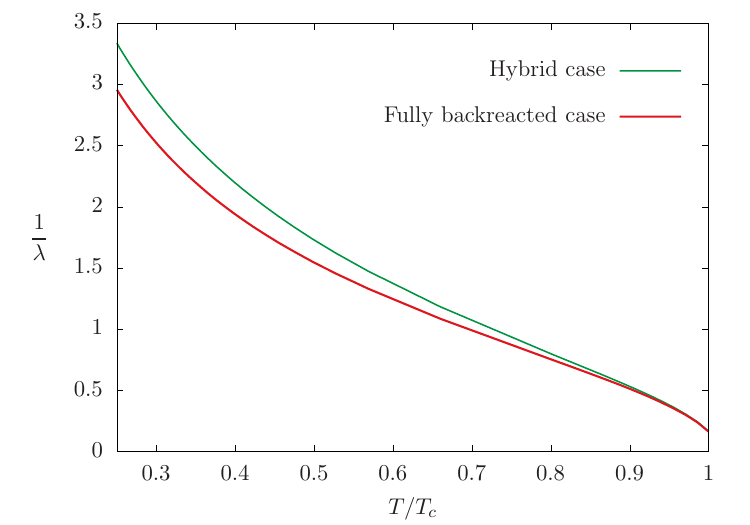}
	\caption{
Plot of $T/T_{c}$ \emph{vs} $\frac{1}{\lambda}$ for fixed $G=0.2$, $b=7$ in the absence of heat flow sources, comparison between the fully backreacted scenario and the hybrid case when computing $\lambda$.  The fact that the green curve of the hybrid case is above the red one of the fully backreacted case indicates that the superconductor of the former is stronger than the one associated with the latter.}
	\label{fig:T-Tc-lambdacomp}
\end{figure}


In Fig. \ref{fig:T-Tc-lambdacomp} we show that in the absence of heat flow sources at the AdS boundary,
the penetration depth at a given temperature is slightly larger in the fully backreacted case compared to the hybrid one. 
This behavior illustrates that additional contributions on the penetration depth arising from the backreacted metric, further ``weaken'' the superconductor's properties.

The penetration depth is deeply related to the superfluid fraction, $\rho_s$, or superfluid density, which is defined as the ratio of the number of superconducting electrons to the total number of electrons. The precise relation is,

\begin{equation}
\rho_{s}(T) = \left( \frac{\lambda(0)}{\lambda(T)} \right) ^{2}.
\label{eq:ps2Fluid}
\end{equation}

This equation allows us to use (\ref{eq:mubetaIbackreaction}) to study the dependence of
$\rho_s$ on the reduced temperature, $t = T/T_c$. The superfluid density plays a central role in the study of high-temperature superconductors, as it is directly related to the system's stiffness against phase fluctuations. In some high-temperature superconductors, particularly those with doping, phase fluctuations are the dominant factor determining the critical temperature \cite{Emery_1995,li2021superconductormetaltransitionoverdoped}.

The explicit expression for $\rho_{s}(t)$ is model-dependent. In the two-fluid model, the following empirical relation is assumed,
\begin{equation}
\label{fig11}
\rho_{s}= 1 - t^{p}.
\end{equation}
Although this expression seems to lack rigorous microscopic justification, it qualitatively produces good results and coincides with the one predicted by BCS theory at low temperatures. By fitting (\ref{fig11}) to results obtained from BCS, the values $p = 2$ for $s$-wave superconductors and $p = 4/3$ for $d$-wave superconductors have been found \cite{Prozorov_2006}.

\begin{figure}
	\centering
	\includegraphics[scale=0.7]{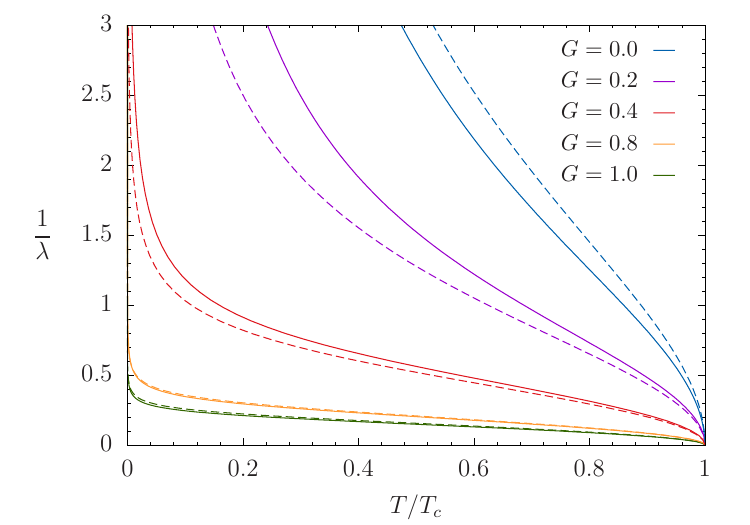}
	\caption{Plot of $T/T_c$ {\it vs} $1/\lambda$ for different values of $b$ and $G$. 
   Dashed curves are for $b=3$, while the continuous one are for $b=7$. The values of $G$ considered are, from top to bottom, $G=0,0.2,0.4,0.8,1.0$. }
	\label{fig:t-lambdaInv}
\end{figure}

In Figure \ref{fig:t-lambdaInv}, we examine the relationship between $\sqrt{\rho_s}$ and $t$. Notice that we have plotted $1/\lambda$, or equivalently $\sqrt{\rho_{s}}$, instead of $\lambda^{-2}$, or equivalently $\rho_{s}$, for better visualization. The concavity of the resulting profiles differs from the typical behaviour found in the superconductivity literature. To our knowledge, such behavior has only been reported in studies on the effects of doping in $d$-wave superconductors (see, e.g., \cite{articleGaniev,osti_1302997,7c2271d3f41f41889ed09beb13b02f28,li2021superconductormetaltransitionoverdoped}). These studies show that increasing doping reduces the superfluid density. This effect is attributed to the correlation length of the disorder introduced by doping. When this correlation length becomes comparable to that of the superconducting condensate, it alters the distribution of the local Cooper pair density, creating regions with high and low pair density \cite{li2021superconductormetaltransitionoverdoped}. This heterogeneity in the superconductor leads to a decrease in both the correlation length and, therefore, the average superfluid density.
Based on this analysis, the non-linearity in the electromagnetic theory of the holographic model appears to capture, to some extent, the experimentally observed doping behaviour, albeit in a spherically symmetric setting.

We want to emphasize that expressions (\ref{eq:lambdaMeissnerProbe}) and
(\ref{eq:mubetaIbackreaction}) are valid only near the critical temperature. To
accurately relate non-linearity to doping-like behavior, terms beyond the second
order in $\psi$ are necessary when computing $\lambda$.
For instance, within this approximation, the values of $p$ and $\lambda(0)$ in Eqs.~(\ref{eq:ps2Fluid})–(\ref{fig11}) cannot be determined. 
A detailed analysis of these aspects is part of an ongoing study.

Moreover, analysing the dependence of these profiles on the parameters $b, G$ and $q$, we see that when the backreaction parameter $G$ is fixed, and we compare the profiles for $b=3$ and $b=7$, approximately below $G \approx 0.1$ and above $G\approx0.8$, the profile for $b=3$ lies above that for $b=7$. In the interval $G\in(0.1, 0.8)$, however, this behavior is inverted, with the $b=3$ profile lying below the $b=7$ one. Nonetheless, at $G=0.1$ and $G=0.8$, both profiles seem to approximately coincide. 
When all the other parameters are fixed and $b$ is decreased ---thereby increasing the degree of non-linearity--- the superfluid density decreases. On the other hand, when  $b$ is fixed and the backreaction parameter $G$ changes, we observe that within the interval $G\in(0.1, 1)$, the superfluid density decreases as $G$ increases. However, for $G=0.03$, this quantity increases compared to its value at $G=0$.

Another aspects worth to mention are the following: when all other parameters are fixed, increasing $q$ ---the coupling constant between the scalar and gauge fields--- has different effects depending on the degree of backreaction. For $G=0.1$, the superfluid density decreases as $q$ increases, as we infer by comparing e.g. the $q=1$ and $q=3$ cases. In contrast, for $G=0.4$, the superfluid density increases when $q$ goes from 1 to 3. However when $q$ is further increased to $5$, the superfluid density decreases compared to the corresponding value at $q=3$. All this information is summarized in Table \ref{Table1}.

\begin{table}[]
\begin{tabular}{ | c c c c  }
\hline
\multicolumn{4}{|c|}{\textbf{Varying $G$} \,  ($b=3,7$ and $q=1$)}    \\ \hline
\multicolumn{4}{|c|}{\quad  $\rho_{G=0.95}$ < $\rho_{G=0.8}$ < $\rho_{G=0.4}$ < $\rho_{G=0.3}$ < $\rho_{G=0.2}$ < $\rho_{G=0.1}$ < $\rho_{G=0}$ < $\rho_{G=0.03}$ \quad \quad }  \\ \hline
\multicolumn{4}{|c|}{\textbf{Varying $b$}  \,  ($G=0.4$ and $q=1$)}    \\ \hline
\multicolumn{4}{|c|}{$\rho_{b=2}$ < $\rho_{b=3}$ < $\rho_{b=7}$ }  \\ \hline
\multicolumn{4}{|c|}{\textbf{Varying $q$}  \,  ($G=0.4$ and $b=3$)}    \\ \hline
\multicolumn{4}{|c|}{$\rho_{q=1}$ < $\rho_{q=5}$ < $\rho_{q=3}$}  \\ \hline
\multicolumn{4}{|c|}{\textbf{Comparison profiles varying $G$}}  \\ \hline
\multicolumn{1}{|c}{\qquad \qquad $G$}  & \multicolumn{1}{c||}{\qquad \quad Profiles ($q$=1) \qquad}   &   \multicolumn{1}{c}{\qquad $G$}  & \multicolumn{1}{c|}{\quad Profiles ($b$=3)}                 \\ 
\multicolumn{1}{|c}{\qquad \qquad 0 -- 0.1}  & \multicolumn{1}{c||}{ \qquad \qquad $\rho_{b=3}$ > $\rho_{b=7}$ \qquad \qquad }  & \multicolumn{1}{c}{}  &   \multicolumn{1}{c|}{}      \\ 
\multicolumn{1}{|c}{ \qquad \qquad 0.1}  & \multicolumn{1}{c||}{\qquad \qquad  $\rho_{b=3}$ $\approx$ $\rho_{b=7}$ \qquad \qquad}  & \multicolumn{1}{c}{\qquad 0.1}  &   \multicolumn{1}{c|}{\quad $\rho_{q=1}$ > $\rho_{q=3}$} \\ 
\multicolumn{1}{|c}{\qquad \qquad 0.2 -- 0.8 \quad}  & \multicolumn{1}{c||}{\qquad \qquad $\rho_{b=3}$ < $\rho_{b=7}$ \qquad \qquad} & \multicolumn{1}{c}{} & \multicolumn{1}{c|}{} \\
\multicolumn{1}{|c}{\qquad \qquad 0.8}  & \multicolumn{1}{c||}{\qquad \qquad  $\rho_{b=3}$ $\approx$ $\rho_{b=7}$ \qquad \qquad} & \multicolumn{1}{c}{\qquad 0.4} & \multicolumn{1}{c|}{\quad $\rho_{q=1}$ < $\rho_{q=3}$} \\
\multicolumn{1}{|c}{ \qquad \qquad 0.8 -- 1}  & \multicolumn{1}{c||}{\qquad \qquad $\rho_{b=3}$ > $\rho_{b=7}$ \qquad \qquad} & \multicolumn{1}{c}{} & \multicolumn{1}{c|}{} \\\hline
\end{tabular}
\caption{Upper lines of the tables show the ordering of superfluid density profiles obtained from varying one parameter, $G, b$ or $q$, while keeping the other two fixed. The lower left corner of the table shows in detail the ordering of the superfluid density profiles for the values $b=3$ and $b=7$, with $q$ fixed, for different values of $G$, whilst the right lower corner compares the superfluid density profiles for $q=1$ and $q=3$, with $b$ fixed, as $G$ varies.}
\label{Table1}
\end{table}

\section{Summary and concluding remarks}
\label{sec:summary}
In this work we have studied numerically a holographic superconductor in a Maxwell-Born-Infeld
setting numerically, taking into account backreaction. Our results coincide qualitatively
with previous results close to the probe limit \cite{Zhao2013,Kruglov2019},
showing that increasing the nonlinearity of the Born-Infeld theory
difficults the production of the superconducting condensate. However,
we found that near the $G=1$ backreacted case, the situation reverses, an increase in the nonlinearity slightly enhances
the production on the condensate. We found a similar behaviour when studying the Meissner effect and the AC conductivity of the system. An important point to be stressed is that backreaction modifies the so-called universal ratio $\omega_g/T_c\sim 8$, making it unstable.

The results reported here are similar to those in \cite{MacedaPatino2019}, where it was seen that in the probe limit,
by introducing a noncommutative spacetime and tuning the respective noncommutativity parameter with the Born-Infeld parameter
in a particular way, the formation of the condensate becomes easier, despite the fact that increasing noncommutativity
tends to difficult the formation of the condensate \cite{Ghorai2016,Pramanik2018}. It is then interesting to
make a numerical study similar to ours for the situation discussed in \cite{MacedaPatino2019} to see how the backreaction changes these
conclusions; this analysis, currently under way, will be reported elsewhere.


To gain more insight into the role played by backreation, we analysed its influence on the penetration depth of the superconductor. Our findings show that for this parameter, when calculated with the so-called superfluid density, the non-linearity of the electrodynamics has a deep impact leading to features similar to doping in superconductors, as reported experimentally in the literature.

 Along these remarks, another interesting line of investigation would be to consider
the Ho\v{r}ava-Lifshitz or other theories of gravity instead of the one of Einstein. In some
settings, it was found that Ho\v{r}ava-Lifshitz gravity difficults the formation of the condensate \cite{Lin2015, Lu2016}
while in others, it does not \cite{Lin2018}. It would be interesting to gain a better understanding of higher derivatives effects, for example, in holographic superconductors and see what happens in the backreacted case.

Finally, we could also study other observables of this superconductor, like its
heat transport or its critical magnetic field and other magnetic properties
\cite{Hartnoll2009,Domenech_2010,Montull_2012}, to pinpoint why nonlinearity
changes from making the condensation harder to easier, when the backreaction
parameter is increased. Such analysis will certainly broaden our knowledge of
strongly correlated superconductors.

\begin{acknowledgments}
  E.G.-G. was funded by a CONAHCYT Scholarship with CVU 494426. U.E. is supported by Mexico’s National Council of Humanities, Science and Technology (CONAHCYT) PhD grant.

\end{acknowledgments}

\section*{Author contributions}
All the authors contributed equally to this article.

\appendix

\section{Conductivity of a Born-Infeld holographic superconductor using the GKPW prescription}
\label{appendix:A}

We recall that for a holographic superconductor based on the Maxwell Lagrangian 
${\cal L}_M = -\frac {1}{16 \pi} F_{\mu\nu} F^{\mu\nu}$,
the relevant quantity used to calculate the conductivity is the following volume integral
that is the contribution in ${\cal L}_M$ that involves $A_x'$
~\cite{Hartnoll2008_Backreaction},
\begin{eqnarray}
-\frac {1}{16\pi} \int_{r_+}^{\infty} dr \int d^3x \, \frac{r^2}{\sqrt n}  \times 
\frac{n}{2 f r^2}(2 f A_x'+g_{xt} \phi')^2,
\end{eqnarray}
which is quadratic on the perturbations $A_x$ and $g_{xt}$ of the background gauge field and metric; the above integral appears in the regularisation procedure of the perturbed hairy black hole action on shell. After
performing
an integration by parts on the variable $r$, together with 
the asymptotic forms in Eq.~(\ref{eq:psiphiInfinityExpansion}) and $A_x = A_x^0 + A_x^1/r +\dots$
to leading order, we get
\begin{equation}
\frac 1{8 \pi}\int d^3x \, A_x^0 A_x^1 \Big|_{r=\infty}+ \dots
\end{equation}
Here the $\dots$ denote contributions independent of the gauge field.
Variation of the above integral with respect to $A_x^0$, together with $A_x^1 \sim A_x^0$, gives 
$J_x \equiv 4\pi (\delta S/\delta A_x^0) = A_x^1$;
in consequence the conductivity is $\sigma \equiv \mathcal J_\mathcal {E}/\mathcal E = -i A_x^1/\omega A_x^0$ ~\cite{Hartnoll2008_Backreaction}.

In the case of Born-Infeld electrodynamics defined by the Lagrangian density
\begin{equation}
{\cal L}_{BI} = 4b^2 \left( 1 - \sqrt{1 + \frac 1{2b^2} F_{\mu\nu} F^{\mu\nu}} \right),
\end{equation}
the corresponding surface integral of interest is 
\begin{eqnarray}
\frac {1}{16\pi} \int_{r_+}^{\infty} dr \int d^3x \, \frac{r^2}{\sqrt n}  \times 
\frac n{2 r^2 f} \frac {(2f A_x' +  g_{tx} \phi')^2}{\sqrt{1 - n b^{-2} \phi'{}^{2}}},
\nonumber\\[4pt] 
\end{eqnarray}
up to second order on the perturbations $A_x$ and $g_{tx}$; the square root in the denominator of the above expression is characteristic of Born-Infeld nonlinear electrodynamics. Following the same procedure as for Maxwell electrodynamics and exploiting the fact that $\phi'|_{r=\infty}=0$ holds as before, a straightforward calculation gives for the previous expression
\begin{equation} 
\frac 1{8\pi} \int d^3x \, A_x^0 A_x^1 \Big|_{r=\infty}+ \dots 
\end{equation}
It follows that we also have $\sigma = -i A_x^1/\omega A_x^0$; this result is expected since far away from a source, Born-Infeld nonlinear electrodynamics behaves similar to Maxwell electrodynamics.

\section{Study of the Meissner effect (hybrid case)}
\label{appendix:B}
To study the penetration depth of a small magnetic field into the superconductor without backreaction, we begin with Eq.~(\ref{eq:meissnerperturb}). Since our focus is on the long-wavelength limit of $A_x$ near the critical temperature,
we propose an expansion around $k=0$ and $\epsilon=0$ of the following form
\begin{equation}
\label{eq:expkeps}
\tilde{A}_{x} = A_{x00} + k^{2} A_{x01} + \epsilon^{2} A_{x10} + \cdots,
\end{equation}
where $\tilde{A}_{x}= \int dq e^{-iky} A_{x}$ is the Fourier transform of $A_x$,
$k$ is the wavelength number and $\epsilon$ is a small parameter of the same order as $\psi$.
This leads us to the following equations
\begin{align}
\label{eq:eqA00}
 \partial_{u} \left( \frac{f \partial_{u} A_{x00}(u)}{\xi(u)} \right) &= 0 \, ,\\
 \partial_{u} \left( \frac{f \partial_{u} A_{x01}(u)}{\xi(u)} \right) - \frac{A_{x00}}{\xi(u)} &= 0 \, , \\
 \partial_{u} \left( \frac{f \partial_{u} A_{x10}(u)}{\xi(u)} \right) -  \frac{8 \pi q^{2} \psi^{2} A_{x00}}{u^{2}}&=0.
\end{align}
To solve the previous system of equations, we first address Eq.~(\ref{eq:eqA00}), which gives the general solution
$A_{x00} \equiv c_{2} J_{A} + c_{1} K_{A} =c_{2} + c_{1} \int_{0}^{1} du' \frac{1}{\xi_{2}(u')} $,
where $\xi (u)\equiv \sqrt{1- u^{4} \phi'^{2}/b^{2}}$ and
$\xi_{2}(u)\equiv f(u)/\xi(u)$. The constants  $c_1$, $c_2$ are  determined
by imposing the boundary conditions $A_{x00}(0)=1$ and $A_{x00}(1) < \infty$, which ensures regularity at the horizon.
These conditions lead to $A_{x00}(u)\equiv1$ identically.

To compute $A_{x01}$ and $A_{x00}$, we will use the machinery of Green's functions.
Consider the Green's function $G_A(u,s)$ for the differential operator in Eq.~(\ref{eq:eqA00}).
By definition, $G_A$ satisfies the partial differential equation
\begin{equation}
 \partial_{u} \left( \frac{f \partial_{u} G_{A}(u,s)}{\xi(u)} \right) = \delta(u-s) \, .
\end{equation}
To ensure that $\tilde A$ is regular at the horizon, and that $\tilde A_x(0)=1$ holds,
we have set the boundary conditions $G_{A}(0,s)=0$, $G_{A}(1,s) < \infty$.
Using the general solution of Eq.~(\ref{eq:eqA00}) found previously, we conclude
\begin{equation}
G_{A}(u,s) = \left\lbrace
\begin{array}{c}
\eta_{1}(s) J_{A} (u) \quad; \quad \quad u > s \\
\eta_{2}(s) K_{A} (u) \quad; \quad \quad u<s
\end{array}
\right. ,
\end{equation}
where $\eta_{1}$ and $\eta_{2}$ are functions determined by imposing $G_{A}(u,s)$ be continuous,
and that its first derivative has the appropriate discontinuity at $s=u$, namely,
\begin{align}
&G_{A}(s + \delta , s)=G_{A}(s - \delta , s) \, , \\
&\int_{s - \delta}^{s + \delta} du \partial_{u} \left[ \xi_{2}(u) \partial_{u} \left( G_{A}(u,s) \right) \right] = 1,
\end{align}
with $\delta \rightarrow 0^+$. From this, we conclude
\begin{align}
G_{A}(u,s) &= \frac{1}{\xi_{2}(s) (K(s)J'(s) - J(s)K'(s))}
\left\lbrace
\begin{array}{c}
K_{A} (s) J_{A} (u) \quad; \quad \quad u > s \\
J_{A} (s) K_{A} (u) \quad; \quad \quad u<s
\end{array}
\right.
\nonumber \\
&= \frac{1}{C_{A}}
\left\lbrace
\begin{array}{c}
K_{A} (s) J_{A} (u) \quad; \quad \quad u > s \\
J_{A} (s) K_{A} (u) \quad; \quad \quad u<s
\end{array}\right. ,
\end{align}
where we have used Abel's formula for the Wronskian of a differential equation to obtain the last expression, namely,
\begin{equation}
\label{eq:wronsk}
W= KJ' - JK' = C_{A} e^{- \int_{0}^{s} du' \frac{\xi_{2}'}{\xi_{2}}} = \frac{C_{A}}{\xi_{2}}.
\end{equation}
Here $C_A$ is a constant.
From the previous analysis, we conclude that $A_{x01}$ and $A_{x10}$ can be written as
\begin{align}
A_{x01}(u)&= \int_{0}^{1} ds \,  G_{A}(u,s) \left( \frac{A_{x00}(s)}{\xi(s)} \right) \, , \\
A_{x10}(u)&= \int_{0}^{1} ds \,  G_{A}(u,s) \left(   \frac{8 \pi q^{2} \psi^{2}(s) A_{x00}(s)}{s^{2}} \right) .
\end{align}
From the standard AdS/CFT dictionary, we know that ${\mathcal{J}}_x = \partial_{u} {A}_{x} \Big\vert_{u=0}$ holds.
By taking the Fourier transform, we obtain $\tilde{\mathcal{J}}_x = \partial_{u} \tilde{A}_{x} \Big\vert_{u=0}$, where
$\tilde{\mathcal{J}}_x$ denotes the Fourier transform of $\mathcal J_x$. Thus, we compute the following terms,
\begin{align}
\partial_{u} A_{x01} \Big\vert_{u=0} &= \frac{1}{C_{A}} \left[ \int_{0}^{u} ds  \left(\frac{K_{A}(s)}{\xi(s)}\right)  \partial_{u} J_{A}(u)   + \int_{u}^{1} ds \left( \frac{J_{A}(s)}{\xi(s)} \right) \left( \partial_{u} K_{A}(u) \right) \right] \Big\vert_{u=0} \nonumber \\
&= \frac{1}{C_{A}} \left[ \int_{0}^{1} ds \left( \frac{1}{\xi_{2}(0) \xi(s)} \right) \right] = \frac{1}{C_{A}} \int_{0}^{1} ds  \frac{1}{\xi(s)},
\end{align}
and
\begin{align}
\partial_{u} A_{x10} \Big\vert_{u=0} &= \frac{1}{C_{A}} \left[ \int_{0}^{u} ds K_{A}(s) \left( \partial_{u} J_{A}(u) \right) + \int_{u}^{1} ds J_{A}(s) \left( \partial_{u} K_{A}(u) \right) \right] \left(  \frac{8 \pi q^{2} \psi^{2}(s)}{s^{2}} \right) \Big\vert_{u=0} \nonumber \\
&= \frac{1}{C_{A}} \left[ \int_{0}^{1} ds \left( \frac{1}{\xi_{2}(0)} \right) \right] \left(  \frac{8 \pi \psi^{2}(s)}{s^{2}} \right) = \frac{1}{C_{A}} \int_{0}^{1} ds \left(  \frac{8 \pi q^{2} \psi^{2}(s)}{s^{2}} \right)\, ,
\end{align}
in which we have used that $A_{x00}\equiv1$.
The value of $C_{A}$ can be obtained using the explicit form of $K_{A}$ and $J_{A}$, along with Eq.~(\ref{eq:wronsk}),
resulting in,
\begin{equation}
C_{A}= \xi_{2} \left( K_{A} J_{A}' - J_{A}K_{A}' \right) = -1.
\end{equation}
By applying these results in Eq.~(\ref{eq:expkeps}),
we conclude that $\tilde{\mathcal J}_x$ can be expressed as,
\begin{align}
\label{eq:mixcond}
& \tilde{\mathcal J}_x = \partial_{u} \tilde{A}_{x} \Big\vert_{u=0} = -k^{2} \beta A_{y00} - 8 \pi q^{2} I A_{y00}\, ,  \\
&\beta= \int_{0}^{1} du \frac{1}{\xi (u)} \, , \quad \quad \quad \quad I= \int_{0}^{1} du \frac{\psi^{2} (u)}{u^{2}}\,.
\end{align}
From these expressions, we can compute the penetration depth.
First, from the mixed boundary condition $\partial_{j} \mathcal{F}^{xj}= 4 \pi \mathcal J_x$,
we obtain,
\begin{equation}
k^{2} \tilde{\mathcal{A}_{x}} = -4 \pi (k^{2} \beta + 8 \pi q^2 I)\tilde{\mathcal{A}_{x}},
\end{equation}
where $\mathcal A_x = A_x |_{u =0}$ denotes the boundary gauge field, and $\tilde{\mathcal A_x}$,
its Fourier transform.
From the above expression, we immediately see that the solution for $\tilde{\mathcal{A}_{x}}$ is
the trivial one. To obtain a non-trivial solution, we follow the approach in
\cite{Natsuume:2022kic} and add an external current $J_x^\text{ext}$.
When we incorporate this term into the mixed boundary conditions, and omit the second term on the r.h.s.\ of Eq.~(\ref{eq:mixcond})
--- the term related to a supercurrent --- we conclude,
\begin{equation}
k^{2} \tilde{\mathcal{A}_{x}} = 4 \pi \mu_{m} J_{x}^\text{ext}, \qquad \mu_{m}= \frac{1}{1+ 4 \pi \beta},
\label{eq:defMu}
\end{equation}
where $\mu_{m}$ denotes the permeability of the normal phase, as can be verified
by noting that,
$\nabla \times \mathcal B = \nabla \left( \nabla \cdot \mathcal{A} \right) - \nabla^{2} \mathcal{A} = k^{2} \tilde{\mathcal{A}_{x}}$,
and by comparing Eq.~(\ref{eq:defMu}) with Ampère's law, $\nabla \times \mathcal B= 4 \pi \mu_m \mathcal J$.

The next step is to include the contribution from the supercurrent present in Eq.~(\ref{eq:mixcond}). The mixed boundary conditions become
\begin{equation}
k^{2} \tilde{\mathcal{A}_{x}} = -4\pi \left( k^{2} \beta + 8 \pi q^{2} I \right) \tilde{A_{x}} + 4\pi J_{x}^\text{ext}\,.
\end{equation}
It follows that
\begin{equation}
\tilde{\mathcal{A}_{x}} = \frac{4 \pi \mu_{m}}{k^{2} + 32 \pi^2 q^{2} \mu_{m} I} J_{x}^\text{ext}.
\end{equation}
From the last expression, we see that
\begin{equation}
\tilde{\mathcal{A}_{x}} \propto \frac{1}{k^{2} + 32 \pi^2 q^{2} \mu_{m} I} \Rightarrow
\mathcal{A}_{x} \propto e^{-x/ \lambda}, \text{ with }\lambda ^{2}\equiv \frac{1}{32 \pi^2 q^{2} \mu_{m} I},
\end{equation}
where we applied the inverse Fourier transform to compute $\mathcal{A}_x$ from
$\tilde{\mathcal{A}_x}$. $\lambda$ is the penetration depth associated with the Meissner
effect.

\section{Study of the Meissner effect with full backreaction}
\label{appendix:C}
Following a similar procedure as in the hybrid case, we compute the penetration depth in the fully backreacted scenario.
Using an expansion in $\epsilon$ and $k$ analogous to that in Eq.~(\ref{eq:expkeps}) for
$\tilde A_x$ and $\tilde g_{tx}$ --- the Fourier transforms of $A_x$ and $g_{tx}$ ---
we obtain the following equations
\begin{align}
\label{eq:Ax00br}
  \partial_{u} \left[ \frac{f \sqrt{n} u^{2} A_{x00}' + \sqrt{n}u^{2} \phi' g_{x00}}{\gamma(u)} \right]&=0\,, \\
  \partial_{u} \left[ \frac{f \sqrt{n} u^{2} A_{x01}' + \sqrt{n}u^{2} \phi' g_{x01}}{\gamma(u)} \right] -  \frac{ A_{x00}}{ \sqrt{n} \gamma(u)}&=0\,, \\
  \partial_{u} \left[ \frac{f \sqrt{n} u^{2} A_{x10}' + \sqrt{n}u^{2} \phi' g_{x10}}{\gamma(u)} \right] - \frac{8 \pi q^{2} \psi^{2}}{f u^{2} \sqrt{n}} \left( f A_{x00} + \phi g_{x00} \right)
  &=0\,,
\end{align}
together with
\begin{align}
\label{eq:gx00br}
 - \sqrt{u^{4}n} \partial_{u} \left( \sqrt{u^{4} n} g_{x00}' \right) + u n^{2} \partial_{u} \left( \frac{u^2}{n} \right) g_{x00} - \frac{4 G u^{4} n \phi' A_{x00}'}{\gamma} &=0 ,\\
 - \sqrt{u^{4}n} \partial_{u} \left( \sqrt{u^{4} n} g_{x01}' \right) + u n^{2} \partial_{u} \left( \frac{u^2}{n} \right) g_{x01} - \frac{4 G u^{4} n \phi' A_{x01}'}{\gamma} + \frac{u^{2} g_{x00}}{f} &= 0 \, , \\
 - \sqrt{u^{4}n} \partial_{u} \left( \sqrt{u^{4} n} g_{x10}' \right) + u n^{2} \partial_{u} \left( \frac{u^2}{n} \right) g_{x10} - \frac{4 G u^{4} n \phi' A_{x10}'}{\gamma} - \frac{32 G \pi q^{2} \phi \psi^{2} A_{x00}}{f} &= 0\,,
\end{align}
where $\gamma(u) \equiv \sqrt{1-u^{4}n \phi'^{2}/b^{2}}$.

As in the hybrid case, we first obtain the general expression for
$A_{x00}$ and use it to define the Green's function that will allow us to calculate
the subsequent orders in the expansion. For this purpose, from Eq.~(\ref{eq:Ax00br}),
we define the differential operator $M$ as
\begin{equation}
 M \left[ A_{x00} \right] = \partial_{u} \left[ \frac{f \sqrt{n} u^{2} A_{x00}'}{\gamma(u)} \right],
\end{equation}
so that Eq.~(\ref{eq:Ax00br}) can be rewritten as
\begin{equation}
M \left[ A_{x00} \right] + \partial_{u} \left[ \frac{ \sqrt{n}u^{2} \phi' g_{x00}}{ \gamma(u)} \right]=0 \,.
\end{equation}
The general solution of the homogeneous part of the previous equation can be expressed as
$A_{M}(u) \equiv  c_{2} J_{M} (u)+ c_{1} K_{M} (u) =c_{2} + c_{1} \int^{u}_{0} \frac{1}{\alpha_{2}(u')} du' $,
where $\alpha_{2}(u) \equiv f(u) \sqrt{n(u)} u^{2}/\gamma(u)$. With this solution, it is
straightforward to write the Green's function for $M$
\begin{align}
G_{M}(u,s) &= \frac{1}{\alpha_{2}(s) (K(s)J'(s) - J(s)K'(s))}
\left\lbrace
\begin{array}{c}
K_{M} (s) J_{M} (u) \quad; \quad \quad u > s \\
J_{M} (s) K_{M} (u) \quad; \quad \quad u<s
\end{array}
\right. \nonumber \\
&= \frac{1}{\alpha_{2}(s) \left( C_{M} e^{- \int_{0}^{s} du' \frac{\alpha_{2}'}{\alpha_{2}}} \right)}
\left\lbrace
\begin{array}{c}
K_{M} (s) J_{M} (u) \quad; \quad \quad u > s \\
J_{M} (s) K_{M} (u) \quad; \quad \quad u<s
\end{array}
\right. \nonumber \\
&= \frac{1}{ C_{M}} \left\lbrace
\begin{array}{c}
K_{M} (s) J_{M} (u) \quad; \quad \quad u > s \\
J_{M} (s) K_{M} (u) \quad; \quad \quad u<s
\end{array}\right. ,
\end{align}
where $C_{M}=-1$, as can be verified using the explicit forms of  $J_{M}$ and
$K_{M}$. With the Green's function $G_{M}$, we can compute the subsequent orders of $\tilde A$ as follows
\begin{align}
\label{eq:Ax00Gr}
A_{x00}&=1- \int_{0}^{1} ds G_{M}(u,s) \left[ \partial_{s} \left( \frac{ \sqrt{n(s)}s^{2} \phi'(s) g_{x00}(s)}{ \gamma(s)} \right) \right]\, , \\
A_{x01}&= \int_{0}^{1} ds G_{M}(u,s) \left[ - \partial_{s} \left( \frac{ \sqrt{n(s)}s^{2} \phi'(s) g_{x01}(s)}{ \gamma(s)} \right) + \frac{A_{x00}}{\sqrt{n(s)} \gamma(s)} \right] \, ,\\
A_{x10}&= \int_{0}^{1} ds G_{M}(u,s) \left[ - \partial_{s} \left( \frac{ \sqrt{n(s)}s^{2} \phi'(s) g_{x10}(s)}{ \gamma(s)} \right) \right. \nonumber \\
\label{eq:Ax10Gr}
& \left.  + \quad \frac{8 \pi q^{2} \psi^{2}(s)}{f(s) s^{2} \sqrt{n(s)}} \left( f(s) A_{x00} +\phi(s) g_{x00} \right) \right],
\end{align}
where the term $c_{2}J_M=1$ in the r.h.s.\ of the first equation was introduced by hand, as we can always add a solution of the homogeneous equation $M[A_{x00}] = 0$, to recover the previous solution $A_{x00} \equiv 1$ in the case where $g_{x00} \equiv 0$.

To further proceed, firstly we compute $A_{x00}$ explicitly, as it is required to compute $A_{x01}$ and $A_{x10}$; we have
\begin{align}
\label{eq:Ax00brfull}
A_{x00}&= 1 - \left\lbrace J_{M}(u) \int_{0}^{u} ds K_{M}'(s) \frac{\sqrt{n(s)} s^{2} \phi'(s) g_{x00}(s)}{ \gamma(s)} \right. \nonumber \\
& \left. \quad + K_{M}(u) \int_{0}^{u} ds J_{M}'(s) \frac{\sqrt{n(s)} s^{2} \phi'(s) g_{x00}(s)}{ \gamma(s)} \right\rbrace \\
&=1 - \int_{0}^{u} ds \frac{\sqrt{n(s)} s^{2} \phi'(s)}{\alpha_{2}(s) \gamma(s)} g_{x00}(s) =1 - \int_{0}^{u} ds \frac{\phi'(s)}{ f(s)} g_{x00}(s)\, .
\end{align}
As in the hybrid case,  computing the penetration depth
requires evaluating the derivatives of Eqs.~(\ref{eq:Ax00Gr})-(\ref{eq:Ax10Gr}) at the boundary, which gives
for $\partial_{u} A_{x00}$ the result
\begin{align}
\partial_{u} A_{x00} \Big\vert_{u=0}&= - \left( \partial_{u} K_{M}(u) \right) \Big\vert_{u=0} \left\lbrace \frac{J_{M}(u) \sqrt{n(u)} \phi'(u) u^{2} g_{x00}(u)}{ \gamma(u)} \Big\vert_{u=0} \right. \nonumber \\
& \left. + \quad \int_{0}^{1} ds \left[ \frac{J_{M}'(s) \sqrt{n(s) \phi'(s) s^{2}} g_{x00}(s)}{ \gamma(s)}  \right] \right\rbrace \\
&= - \frac{\phi'(u) g_{x00}(u)}{f(u)} \Big\vert_{u=0} = \rho g_\text{bdry0} ,
\end{align}
 where we used the asymptotic expansions of Eq.~(\ref{eq:psiphiInfinityExpansion}),
and defined $g_\text{bdry0}$ as the leading-order coefficient in the
expansion of $g_{x00}$ around the boundary, i.e.
$g_{x00} = g_\text{bdry0}/ u^{2} + g_{1} u + \cdots$. Similarly, for $\partial_{u} A_{x01} \Big\vert_{u=0}$,
we find
\begin{align}
\partial_{u} A_{x01} \Big\vert_{u=0}&= - \left( \partial_{u} K_{M}(u) \right) \Big\vert_{u=0} \left\lbrace \frac{J_{M}(u) \sqrt{n(u)} \phi'(u) u^{2} g_{x01}(u)}{\gamma(u)} \Big\vert_{u=0} \right. \nonumber \\
& \left. + \quad \int_{0}^{1} ds \left[ \frac{J_{M}'(s) \sqrt{n(s)} \phi'(s) s^{2} g_{x01}(s)}{\gamma(s)} + \frac{ J_{M}(s) A_{x00}(s)}{\sqrt{n(s)} \gamma(s)} \right] \right\rbrace \nonumber \\
&= - \left\lbrace - \rho  g_{x01}(0)+ \int_{0}^{1} ds \left[ \frac{A_{x00}(s)}{\sqrt{n(s)} \gamma(s)} \right] \right\rbrace \nonumber \\
\label{20}
&= \rho  g_{x01}(0) - \int_{0}^{1} ds \left[ \frac{A_{x00}(s)}{\sqrt{n(s)} \gamma(s)} \right],
\end{align}
and for $\partial_{u} A_{x10} \Big\vert_{u=0}$ we get
\begin{align}
\partial_{u} A_{x10} \Big\vert_{u=0}&= - \left( \partial_{u} K_{M}(u) \right) \Big\vert_{u=0} \left\lbrace \frac{J_{M}(u) \sqrt{n(u)} \phi'(u) u^{2} g_{x10}(u)}{ \gamma(u)} \Big\vert_{u=0} + \int_{0}^{1} ds \left[ \frac{J_{M}'(s) \sqrt{n(s)} \phi'(s) s^{2} g_{x10}(s)}{ \gamma(s)} \right. \right. \nonumber \\
& \left. \left. \quad + \frac{8 \pi q^{2} \psi^{2}(s) J_{M}(s)}{f(s) s^{2} \sqrt{n(s)}} \left(  f(s) A_{x00}(s) + \phi(s) g_{x00}(s) \right) \right] \right\rbrace \nonumber \\
&= - \left\lbrace - \rho  g_{x10}(0) + \int_{0}^{1} ds \left[ \frac{8 \pi q^{2} \psi^{2}(s)}{f(s) s^{2} \sqrt{n(s)}} \left(  f(s) A_{x00}(s) + \phi(s) g_{x00}(s) \right) \right] \right\rbrace \nonumber \\
=& \rho  g_{x10}(0) -  8 \pi q^{2}\int_{0}^{1} ds \left[ \frac{ \psi^{2}(s)}{f(s) s^{2} \sqrt{n(s)}} \left(  f(s) A_{x00}(s) +\phi(s) g_{x00}(s)\right) \right]\,.
\label{21}
\end{align}
Although it may initially appear that the full profiles of $g_{x01}$ and
$g_{x10}$ are needed to compute the penetration depth, the previous equations
reveal that only their leading-order coefficients near the boundary,
$g_{x01}(0)$ and $g_{x10}(0)$, are actually required.
If we assume that $g_\text{bdry0}$ is fixed to a certain value independently of the condensate and the wavelength number
(which, in the context of holography corresponds to a fixed source of heat flow at the AdS boundary \cite{Hartnoll2008_Backreaction}),
we then have $g_{x01}(0)=g_{x10}(0)=0$.

After imposing the mixed boundary conditions we obtain
\begin{align}
&\qquad \qquad \qquad k^{2} \tilde{\mathcal{A}_{x}} = -4 \pi \left[ \theta_{1}+ k^{2}   \beta''  + 8 \pi q^{2}   I''  \right] \tilde{\mathcal{A}_{x}}, \\
&\theta_{1}= -\partial_{u} A_{x00} \Big\vert_{u=0}, \quad   \beta'' = -\partial_{u} A_{x01} \Big\vert_{u=0},  \quad   8 \pi q^2 I''= -\partial_{u} A_{x10} \Big\vert_{u=0}.
\end{align}
Using the same reasoning as in Appendix~\ref{appendix:B}, after introducing an external current,
we can derive the penetration depth for the backreacted case
\begin{equation}
 \lambda ^{2}= \frac{1}{ 4 \pi \mu_{m} \left( \theta_{1} +  8 \pi q^{2} I'' \right)},
\qquad \mu_{m}= \frac{1}{1+ 4\pi \beta''}.
\label{eq:lamba2Final}
\end{equation}
Finally, we address the numerical solution of the necessary equations.
By solving for $A_{x00}'$ from Eq.~(\ref{eq:gx00br}) and substituting it in Eq.~(\ref{eq:Ax00br}), we derive the following
differential equation
\begin{align}
\label{eq:JKg}
\partial_{u} \left[ \frac{f \sqrt{n}}{4 G \phi'}  \left( -  \frac{1}{\sqrt{n}} \partial_{u} \left( \sqrt{u^{4} n} g_{x00}' \right) + \frac{n}{u} \partial_{u} \left( \frac{u^2}{n} \right) g_{x00} \right) + \frac{\sqrt{n}u^{2} \phi' g_{x00}}{\gamma} \right]=0 .
\end{align}
To impose regularity of the solution, $g_{x00}$ must vanish at the horizon \cite{Hartnoll2008_Backreaction}. Near the boundary,
$g_{x00}$ behaves as  $g_{x00} \approx g_\text{bdry0}/u^2$. By fixing $g_\text{bdry0}$ to a particular value, we have established the
boundary conditions required to compute $g_{x00}$ numerically.
Once $g_{x00}$ is computed, $A_{x00}$ can be obtained from Eq.~(\ref{eq:Ax00brfull}), and from there, we can compute Eqs.~(\ref{20}) and (\ref{21})
numerically.
When there are no sources of heat flow, \emph{i.e.}\ $g_\text{bdry0}=0$, many expressions simplify considerably, as
this condition implies that $g_{x00}\equiv0$, which in turn gives  $A_{x00} \equiv 1$. Thus, the expressions in Eq.~(\ref{eq:mubetaIbackreaction}) in the main text
are established.

\bibliography{draft.bib}

\end{document}